\documentclass[reqno,a4paper,11pt]{article}
\pdfoutput=1
\usepackage{color}

\usepackage{graphicx}
\usepackage[textwidth = 430 pt, textheight = 630 pt]{geometry}

\definecolor{MyDarkBlue}{rgb}{0.15,0.25,0.45}
\usepackage{epsfig,rotating}
\usepackage{amsmath,amssymb}
\usepackage{amsfonts}
\usepackage{mathrsfs}
\usepackage{bbm}
\usepackage[normalem]{ulem}

\usepackage{latexsym}
\usepackage{amsthm}

\usepackage[utf8x]{inputenc}

\usepackage{hyperref}
\hypersetup{
hypertexnames=false,
colorlinks=true,
citecolor=MyDarkBlue,
linkcolor=MyDarkBlue,
urlcolor=MyDarkBlue,
pdfauthor={Christian S\"amann},
pdftitle={Bootstrapping fuzzy scalar field theory},
pdfsubject={hep-th math-ph}
breaklinks=true
}

\usepackage{tikz}
\usepackage{mathtools}

\let\fn\footnote
\renewcommand{\footnote}[1]{\linespread{1.1}\fn{#1}\linespread{1.29}}

\makeatletter\renewcommand{\section}{\@startsection
{section}{1}{\z@}{-3.5ex plus -1ex minus
    -.2ex}{2.3ex plus .2ex}{\bf }}
\makeatletter\renewcommand{\subsection}{\@startsection{subsection}{2}{\z@}{-3.25ex
plus -1ex minus
   -.2ex}{1.5ex plus .2ex}{\it }}
\makeatletter\renewcommand{\subsubsection}{\@startsection{subsubsection}{3}{-2.45ex}{-3.25ex
plus -1ex minus -.2ex}{1.5ex plus .2ex}{\it }}
\renewcommand{\thesection}{\arabic{section}}
\renewcommand{\thesubsection}{\arabic{section}.\arabic{subsection}}
\renewcommand{\@seccntformat}[1]{\@nameuse{the#1}.~~}

\renewcommand{\theequation}{\thesection.\arabic{equation}}
\makeatletter \@addtoreset{equation}{section}

\usepackage[toc,page]{appendix}

\renewcommand{\thethm}{\thesection.\arabic{thm}}

\renewcommand{\appendices}{
\section*{Appendix}\label{appendices}\setcounter{subsection}{0}
\addcontentsline{toc}{section}{Appendix}
\setcounter{equation}{0}
\makeatletter
\renewcommand{\theequation}{\Alph{subsection}.\arabic{equation}}
\renewcommand{\thesubsection}{\Alph{subsection}}
\renewcommand{\thethm}{\Alph{subsection}.\arabic{thm}}
\@addtoreset{equation}{subsection}
\@addtoreset{thm}{subsection}
\makeatother
}

\usepackage[all,knot]{xy}
\xyoption{arc}

\def\slasha#1{\setbox0=\hbox{$#1$}#1\hskip-\wd0\hbox to\wd0{\hss\sl/\/\hss}}

\def\periodb#1{\setbox0=\hbox{$#1$}#1\hskip-\wd0\hbox to\wd0{-}}

\newcommand{\unit}{\mathbbm{1}}   			

\newcommand{\CCD}{\mathscr{D}}

\newcommand{\CH}{\mathcal{H}}

\newcommand{\CO}{\mathcal{O}}

\newcommand{\CZ}{\mathcal{Z}}

\newcommand{\FR}{\mathbbm{R}}     			
\newcommand{\FC}{\mathbbm{C}}     			
\newcommand{\NN}{\mathbbm{N}}     			
\newcommand{\CPP}{{\mathbbm{C}P}}    			

\newcommand{\dd}{\mathrm{d}}     			
\newcommand{\dpar}{\partial}     			
\newcommand{\de}{\mathrm{e}}     			
\newcommand{\di}{\mathrm{i}}     			
\newcommand{\eps}{{\varepsilon}}			

\newcommand{\eand}{{\qquad\mbox{and}\qquad}}     		

\newcommand{\der}[1]{\frac{\dpar}{\dpar #1}}   		
\newcommand{\tr}{\,\mathrm{tr}\,}     			

\newcommand{\au}{\mathfrak{u}}
\newcommand{\asu}{\mathfrak{su}}

\newcommand{\sU}{\mathsf{U}}     			

\newcommand{\sSU}{\mathsf{SU}}

\newcommand{\sEnd}{\mathsf{End}\,}

\newcommand{\acton}{\vartriangleright}     			
\def\tyng(#1){\hbox{\tiny$\yng(#1)$}}			
\def\tyoung(#1){\hbox{\tiny$\young(#1)$}}			

\newcommand{\beq}{\begin{eqnarray}}
\newcommand{\eeq}{\end{eqnarray}}

\begin{document}

\begin{titlepage}
\begin{flushright}
 EMPG--14--23 
\end{flushright}
\vskip 2.0cm
\begin{center}
{\LARGE \bf Bootstrapping Fuzzy Scalar Field Theory}
\vskip 1.5cm
{\Large Christian S\"amann}
\setcounter{footnote}{0}
\renewcommand{\thefootnote}{\arabic{thefootnote}}
\vskip 1cm
{\em Maxwell Institute for Mathematical Sciences\\
Department of Mathematics, Heriot-Watt University\\
Colin Maclaurin Building, Riccarton, Edinburgh EH14 4AS, U.K.}\\[0.5cm]
{Email: {\ttfamily c.saemann@hw.ac.uk}}
\end{center}
\vskip 1.0cm
\begin{center}
{\bf Abstract}
\end{center}
\begin{quote}
We describe a new way of rewriting the partition function of scalar field theory on fuzzy complex projective spaces as a solvable multitrace matrix model. This model is given as a perturbative high-temperature expansion. At each order, we present an explicit analytic expression for most of the arising terms; the remaining terms are computed explicitly up to fourth order. The method presented here can be applied to any model of hermitian matrices. Our results confirm constraints previously derived for the multitrace matrix model by Polychronakos. A further implicit expectation about the shape of the multitrace terms is however shown not to be true.
\end{quote}
\end{titlepage}

\section{Introduction}

Fuzzy complex projective spaces provide an interesting and highly symmetric way of regularizing quantum field theories in even dimensions \cite{Madore:1991bw,Grosse:1995ar}. These quantum spaces are captured by a finite dimensional matrix algebra of functions labeled by an integer $\ell$. Taking the limit $\ell\rightarrow \infty$, one recovers the ordinary algebra of functions on complex projective space in a controlled manner. A real scalar field on fuzzy $\CPP^n$ is simply a finite-dimensional hermitian matrix and scalar field theory on fuzzy $\CPP^n$ therefore reduces to a hermitian matrix model. We will refer to these models collectively as fuzzy scalar field theories. Note that the partition function of the latter is automatically well-defined for positive actions, as only a finite number of integrals have to be performed. Thus, a possible regularization prescription for quantum scalar field theories in even dimensions would be to put them on fuzzy complex projective space and to take a large $\ell$ limit.

It is therefore interesting to study fuzzy scalar field theories in detail and to try to obtain a good handle on them. In the past, these theories have been studied extensively using numerical methods, see e.g.\ \cite{Martin:2004un,GarciaFlores:2005xc,Panero:2006bx,GarciaFlores:2009hf,Ydri:2014rea}. A first analytical study of fuzzy scalar field theories was performed in \cite{Steinacker:2005wj} for the limit of large matrix sizes. 

Although fuzzy scalar field theories are very close to well-studied hermitian matrix models, they crucially differ in the appearance of fixed matrices in their kinetic terms. These fixed matrices present an obstacle to applying the standard methods for solving hermitian matrix models. To overcome this problem, one can rewrite the kinetic terms as trace and multitrace expressions, and a perturbative way of doing this was given in \cite{O'Connor:2007ea}. This perturbative rewriting essentially corresponds to a high-temperature expansion combined with group-theoretic methods. The rewritten model can then be treated by the standard methods of matrix models. For example, in the limit of large matrix sizes, a saddle point approximation can be used to study the phase diagram of the model. This was done to quadratic order in the inverse temperature for the fuzzy sphere in \cite{O'Connor:2007ea} and to cubic order and for all fuzzy complex projective spaces in \cite{Saemann:2010bw}. 

The same method has been subsequently used in \cite{Ihl:2010bu} to study scalar field theories on a space-time consisting of the real line times a fuzzy complex projective space. Moreover, it was used in various other contexts as e.g.\ \cite{Polychronakos:2013nca,Tekel:2014bta,Ydri:2014uaa}. Particularly interesting are statements about the shape of the multitrace rewriting in the large $N$ limit for the fuzzy sphere made in \cite{Polychronakos:2013nca}. Since the eigenvalue density of free scalar quantum field theory is a renormalized Wigner semicircle as shown in \cite{Steinacker:2005wj}, see also \cite{Nair:2011ux}, many terms in the multitrace rewriting are not allowed to survive the large $N$ limit. In \cite{Polychronakos:2013nca}, this result was used to give an all-order form for the rewriting modulo some remainder terms $R$:
\begin{equation}\label{eq:conjectured_action}
 S_{\rm MT,kin}=F\left(\tr(\Phi^2)-\frac{1}{N}\tr(\Phi)^2\right)+R~,~~~F(t)=N^2\ln\frac{t}{1-\de^{-t}}~.
\end{equation}
Here, $\Phi$ denotes the hermitian $N\times N$ matrix capturing the scalar field. It was implicitly hoped in \cite{Polychronakos:2013nca} that $R$ vanishes, at least in the large $N$ limit.

A first motivation for the present work is to verify this expectation. If true, this would also suggest that there might be a complete analytical description of fuzzy scalar field theory. The striking closeness of fuzzy scalar field theory to integrable matrix models together with its large amount of symmetry makes it hard to believe that there is no such description. Further motivation stems from the vast number of noncommutative field theories that are essentially hermitian matrix models and could benefit from strong techniques to formulate high-temperature expansions.

This paper is structured as follows. In section 2 we briefly review fuzzy scalar field theory and what is known about its high-temperature expansion, fixing our conventions. We then describe in detail a bootstrapping method that yields sufficiently many constraints on the exponentiated action to fix the high-temperature expansion in section 3. The results are then combined in section 4, where the full multitrace expression for the kinetic term is given to fourth order in the inverse temperature $\beta$. There, we also study the large $N$ limit and compare our results to those of \cite{Polychronakos:2013nca,Tekel:2014bta,Nair:2011ux}. We conclude in section 5. A concise summary of our conventions is given in an appendix.

\section{Fuzzy scalar field theory}

\subsection{Fuzzy complex projective spaces}

The fuzzy sphere is the geometric or rather Berezin quantization of the K\"ahler manifold $\CPP^1$, see \cite{Berezin:1974du,Hoppe:Diss,Madore:1991bw}. The Hilbert space $\CH_\ell$ at level $\ell\in \NN_0$ in this quantization is identified with the global holomorphic sections of the $\ell$-fold tensor product of $\CO(1)$, the dual of the tautological line bundle over $\CPP^1$. We thus have $N:=\dim(\CH_\ell)=\ell+1$. The hermitian elements of the vector space of endomorphisms $\sEnd(\CH_\ell)$ of this Hilbert space are in one-to-one correspondence with the spherical harmonics truncated at angular momentum $\ell$ or, equivalently, at eigenvalue $2\ell(\ell+1)$ of the Laplace operator. The product between hermitian elements of $\sEnd(\CH_\ell)$ corresponds to a deformation of the product of the corresponding spherical harmonics of order $\ell^{-1}$. The Hilbert space $\CH_\ell$ carries a representation $\rho_\ell$ of $\sSU(2)$, and thus $\sEnd(\CH_\ell)$ carries the representation $\bar\rho_\ell\otimes \rho_\ell$. The Laplace operator on the fuzzy sphere is identified with the second Casimir acting on the latter representation. Its eigenvalues are $2k(k+n)$ with multiplicities $2k+1$, where $k=0,\ldots,\ell$ .

Fuzzy complex projective spaces are obtained as a straightforward generalization of the fuzzy sphere. In particular, the Berezin quantization of the K\"ahler manifold $\CPP^n$ yields a Hilbert space $\CH_\ell\cong H^0(\CO(1)^{\otimes\ell})$ with $N=\dim(\CH_\ell)=\frac{(n+\ell)!}{n!\ell!}$. Again, the algebra $\sEnd(\CH_\ell)$ approximates a complexification of the algebra of functions on $\CPP^n$, truncated at eigenvalue $2\ell(\ell+n)$ of the Laplace operator. Taking the limit $\ell\rightarrow \infty$, the functions on $\CPP^n$ are recovered in a controlled way. The Hilbert space $\CH_\ell$ now carries a representation $\rho$ of $\sSU(n+1)$ and we identify the Laplace operator on fuzzy $\CPP^n$ with the second Casimir acting on the representations $\bar\rho_\ell\otimes \rho_\ell$ formed by $\sEnd(\CH_\ell)$. Its eigenvalues are $2k(k+n)$ with multiplicities $\frac{n(2k+n)((k+n-1)!)^2}{(k!)^2(n!)^2}$, where again $k=0,\ldots,\ell$. 

For more details on the fuzzy geometry of $\CPP^n$, see \cite{O'Connor:2007ea,Saemann:2010bw} and references therein.

\subsection{The model}

As stated in the introduction, one of the key reasons for studying fuzzy spaces is their capability of regularizing Euclidean path integrals. Since the function algebra is finite dimensional, the Euclidean path integral for scalar fields consists of a finite number of ordinary integrals and there is a controlled limiting procedure that turns the finite dimensional matrix algebra into the algebra of functions on a complex projective space.

Above, we collected all the necessary details to write down an action for scalar field theory on fuzzy $\CPP^n$. For a quantization at level $\ell$ with Hilbert space $\CH_\ell$ of dimension $N$, a scalar field is an element of $\sEnd(\FC^N)$. For a real such field, which is encoded in a hermitian $N\times N$-matrix $\Phi$, we define the functional
\begin{equation}\label{eq:original_action}
 S[\Phi]:=\tr(\Phi~C_2\acton \Phi+r\Phi^2+g\Phi^4)=\tr(\Phi [L_i,[L_i,\Phi]]+r\Phi^2+g\Phi^4)~.
\end{equation}
Here, the $L_i$, $i=1,\ldots,(n+1)^2-1$, are $N\times N$-matrices, forming an $N$-dimensional representation of the algebra $\asu(n+1)$ of isometries of $\CPP^n$:
\begin{equation}
 [L_i,L_j]=:\di f_{ijk}L_k~.
\end{equation}
The generators $L_i$ are normalized such that the eigenvalues of $C_2$ are $2k(k+n)$, as fixed above.

The corresponding Euclidean path integral or partition function is
\begin{equation}\label{eq:partition_function}
 \CZ:=\int \CCD\Phi~\de^{-\beta S[\Phi]}:=\int \dd\mu_D(\Phi)~\de^{-\beta S[\Phi]}~,
\end{equation}
where $\dd\mu_D(\Phi)$ is the usual Dyson measure on the space of hermitian matrices of dimension $N\times N$. 

Note that $\CZ$ is simply the partition function of a hermitian matrix model. To solve such a model, one usually diagonalizes the matrix $\Phi$ as $\Phi=\Omega^\dagger \Lambda \Omega$, where $\Omega$ is a unitary matrix and $\Lambda$ is a diagonal matrix containing the eigenvalues of $\Phi$. The Dyson measure splits accordingly into integrals over eigenvalues and the Haar measure $\dd\mu_H(\Omega)$ on $\sU(N)$:
\begin{equation}
 \int \dd\mu_D(\Phi)=\int\prod_{a=1}^N\dd{\lambda_a}~\Delta^2(\Lambda)\int\dd \mu_H(\Omega)~.
\end{equation}
The Jacobian of this coordinate change gives rise to the square of $\Delta(\Lambda)$, the Vandermonde determinant
\begin{equation}
 \Delta(\Lambda)\ :=\ \det ([\lambda_a^{b-1}]_{ab})\ =\ \prod_{a>b} (\lambda_a-\lambda_b)~.
\end{equation}
The most commonly studied hermitian matrix models do not depend on the angular variables contained in $\Omega$ and the integral over the Haar measure is trivially performed. After integrating out these zero modes, one can apply various techniques to solve the model and to compute the partition function.

Our model \eqref{eq:partition_function} contains a set of $(n+1)^2-1$ ``external matrices'' $L_i$, which present an obstacle to integrating over the Haar measure. Various techniques have been developed for overcoming such problems in certain special cases, but these do not include our particular situation here.

To make progress with the integration over the Haar measure, we first note that the expected result is a function of the eigenvalues of $\Phi$ which is invariant under permutations of these eigenvalues. All such functions of eigenvalues are obtained by integrating the exponential of action functionals of $\Phi$ built from traces of powers of matrices over the Haar measure. We can thus reformulate our action functional as a multitrace matrix model, which has the same partition function and the same expectation values of operators invariant under $\Phi\rightarrow \Omega^\dagger \Phi \Omega$. Moreover, the action of the multitrace matrix model will merely consist of terms containing an even total power of $\Phi$,
\begin{equation}
 S_{\rm MT}[\Phi]=a_2\tr(\Phi^2)+a_{1,1}\tr(\Phi)\tr(\Phi)+\ldots~,
\end{equation}
because our model is invariant under $\Phi\rightarrow -\Phi$. Since products of traces with total power $\alpha$ of $\Phi$ are in one-to-one correspondence to partitions of $\alpha$, we expect one term for any partition of an even integer $\alpha$. It remains to find the corresponding coefficients.

\subsection{Iterative solution via high-temperature expansion}

The observation that quantum scalar field theory on the fuzzy sphere can be reformulated as a multitrace matrix model motivated the constructions in \cite{O'Connor:2007ea,Saemann:2010bw}. While a direct integration over the Haar measure is not possible, the problematic term, i.e.\ the exponential of the kinetic part of the action, can be dealt with in other ways. Its perturbative expansion around 1, which is essentially a high-temperature expansion, reads as 
\begin{equation}
 \de^{-\beta \tr(\Phi~C_2\acton \Phi)}=1-\beta \tr(\Phi~C_2\acton \Phi)+\frac{\beta^2}{2}\left(\tr(\Phi~C_2\acton \Phi)\right)^2+\ldots
\end{equation}
The products of traces can be rewritten as single traces over tensor products. Decomposing these tensor products into irreducible representations, we can use the orthogonality relation of the Haar measure to perform the integral over $\Omega$ order by order in the above expansion. The result can be rephrased as multitrace expressions which, after re-exponentiation, yield the desired multitrace matrix model. In \cite{O'Connor:2007ea}, this method was used to calculate the multitrace matrix model for the fuzzy sphere up to order $\beta^2$. This result was then extended to order $\beta^3$ and generalized to arbitrary fuzzy complex projective spaces in \cite{Saemann:2010bw}. Explicitly, it was found that the multitrace action reads as
\begin{equation}
\begin{aligned}
 S_{\rm MT}[\Phi]&=S_{\rm MT,kin}[\Phi]+\tr(r\Phi^2+g\Phi^4)\\
 &=\frac{\Sigma_1}{N^2-1}\left(\tr(\Phi^2)-\frac{1}{N}\tr(\Phi)^2\right)+\ldots+\tr(r\Phi^2+g\Phi^4)~,
\end{aligned}
\end{equation}
where $\Sigma_1$ is the sum over all eigenvalues of $C_2$ and $\ldots$ denotes multitrace terms of higher orders in $\beta$.

\section{Bootstrapping the model}

\subsection{Basic idea}

Instead of performing the high-temperature expansion as in \cite{O'Connor:2007ea,Saemann:2010bw}, which involved rather tedious group-theoretic computations, we will determine the multitrace action by deriving a number of conditions, which are sufficient to fix it.

Underlying our method is an observation that is also used in some of the proofs of the Harish-Chandra integral formula. Consider some differential operator $D$ such that 
\begin{equation}
 D \de^{-\beta S[\Phi]}=O[\Phi]\de^{-\beta S[\Phi]}~,
\end{equation}
where $O[\Phi]$ is some functional of $\Phi$ invariant under the transformation $\Phi\rightarrow \Omega^\dagger \Phi \Omega$ for unitary matrices $\Omega$. It follows that
\begin{equation}
 D \int \dd \mu_\Omega~\de^{-\beta S[\Omega^\dagger\Phi\Omega]}=O[\Phi]\int \dd \mu_\Omega~\de^{-\beta S[\Omega^\dagger\Phi\Omega]}~.
\end{equation}
Since multitrace expressions are invariant under $\Phi\rightarrow \Omega^\dagger \Phi \Omega$, we can also conclude the same relation for the exponential of the multitrace action $S_{\rm MT}[\Phi]$:
\begin{equation}
 D \de^{-\beta S_{\rm MT}[\Phi]}=O[\Phi]\de^{-\beta S_{\rm MT}[\Phi]}~.
\end{equation}
This relation constrains $S_{\rm MT}[\Phi]$ and, given sufficiently many differential operators $D$, fixes it completely. As we shall see, the result turns out to be a perturbative series in the inverse temperature $\beta$, where the order in $\beta$ is half the total order in the fields $\Phi$.

Since the potential part of the action is already trivially in a multitrace form, we can restrict our attention to the kinetic term, yielding the kinetic part of the multitrace action, $S_{\rm MT,kin}[\Phi]$.

\subsection{Symmetries}

When looking for suitable differential operators to fix the multitrace action $S_{\rm MT, kin}[\Phi]$, we should clearly start by considering the symmetries of the kinetic term $\tr(\Phi~C_2\acton \Phi)$. At infinitesimal level, the rotational symmetries of the sphere act as $\Phi\rightarrow \Phi+\eps^i[L_i,\Phi]$. However, all multitrace actions are invariant under these, so there is nothing to be learnt from them.

Another symmetry is the invariance of the kinetic term under shifts by constant functions. On fuzzy complex projective space, this amounts to an invariance under $\Phi\rightarrow \Phi+c\unit$ for any $c\in \FR$, which is generated by $\der{\Phi_{aa}}:=\sum_a\der{\Phi_{aa}}$. We readily compute
\begin{equation}\label{eq:result_trace}
 \der{\Phi_{aa}}\de^{-\beta\tr(\Phi[L_i,[L_i,\Phi]])}
 =\left((-2\beta\tr([L_i,[L_i,\Phi]]))\de^{-\beta\tr(\Phi[L_i,[L_i,\Phi]])}\right)=0~.
\end{equation}
Therefore, we also conclude that
\begin{equation}\label{eq:condition_trace}
 \der{\Phi_{aa}}\de^{-\beta S_{\rm MT,kin}[\Phi]}=0~.
\end{equation}
Consider the terms in $S_{\rm MT,kin}[\Phi]$ of total power $\alpha$ in $\Phi$. There are $p(\alpha)$ of these, where $p(\alpha)$ gives the number of integer partitions of $\alpha$, and thus there are $p(\alpha)$ coefficients to be fixed. We shall label these coefficients by $a_{\pi_1,\pi_2,\ldots,\pi_k}$, where $\pi_1+\pi_2+\ldots+\pi_k$ is a partition of $\alpha$:
\begin{equation}
\begin{aligned}
 S_{\rm MT}[\Phi]=&a_2\tr(\Phi^2)+a_{1,1}\tr(\Phi)\tr(\Phi)+a_4\tr(\Phi^4)+a_{3,1}\tr(\Phi^3)\tr(\Phi)+\\
 &+a_{2,2}\tr(\Phi^2)^2+a_{2,1,1}\tr(\Phi^2)\tr(\Phi)^2+a_{1,1,1,1}\tr(\Phi)^4+\ldots~.
\end{aligned}
\end{equation}
Equation \eqref{eq:condition_trace} yields conditions $\CO[\Phi]=0$, where $\CO[\Phi]$ consists of multitrace expressions of total power $\alpha-1$ in $\Phi$. Since multitraces are linearly independent for sufficiently large $N$, each of these terms have to vanish separately. This yields $p(\alpha-1)$ conditions on these terms which we can use to express all coefficients of the form $a_{\pi_1,\ldots,\pi_{k-1},1}$ in terms of other coefficients. In particular, consider the multitrace term in $\CO[\Phi]$ corresponding to the partition $\pi_1+\ldots+\pi_{k-1}=\alpha-1$. In terms of coefficients appearing in $S_{\rm MT,kin}$, its vanishing amounts to
\begin{equation}\label{eq:first_fixing}
\begin{aligned}
 \alpha_{\pi_1,\pi_2,\ldots,\pi_{k-1},1}=-\frac{1}{rN}\sum_{\sigma}\Big(&(\sigma(\pi_1)+1)a_{\sigma(\pi_1)+1,\sigma(\pi_2),\ldots,\sigma(\pi_{k-1})}+\\
 &+(\sigma(\pi_2)+1)a_{\sigma(\pi_1),\sigma(\pi_2)+1,\ldots,\sigma(\pi_{k-1})}+\ldots\\
 &\hspace{1cm}+(\sigma(\pi_{k-1})+1)a_{\sigma(\pi_1),\sigma(\pi_2),\ldots,\sigma(\pi_{k-1})+1}\Big)~,
\end{aligned}
\end{equation}
where the sum runs over all permutations of $\pi_1,\ldots,\pi_{k-1}$ and $r-1$ is the number of parts $\pi_i$ which are 1. Moreover, we define $a_{\pi_1,\pi_2,\ldots,\pi_{k-1}}:=0$ unless $\pi_1\geq\pi_2\geq\ldots\geq\pi_{k-1}$. We have for example
\begin{equation}
\begin{aligned}
 a_{1,1}&=-\frac{1}{2N}~2a_2~,~~~&a_{3,1}&=-\frac{1}{N}~4 a_4~,\\
 a_{1,1,1,1}&=-\frac{1}{4N}~2a_{2,1,1}~,~~~&a_{2,1,1}&=-\frac{1}{2N}(3 a_{3,1}+2\cdot 2\cdot a_{2,2})~,\\
 a_{1,1,1,1,1,1}&=-\frac{1}{6N}~2a_{2,1,1,1,1}~,~~~&a_{4,1,1}&=-\frac{1}{2N}(5 a_{5,1}+2 a_{4,2})~. 
\end{aligned}
\end{equation}
Since $p(\alpha-1)\geq \tfrac12 p(\alpha)$, equation \eqref{eq:first_fixing} fixes more than half the unknown coefficients appearing in $S_{\rm MT,kin}[\Phi]$. We shall see below that the coefficients $a_{\pi_1}$ and $a_{\pi_2}$, where $\pi_1$ and $\pi_2$ are partitions of the same integer, scale equally with $\beta$. This is reflected in the linearity of formula \eqref{eq:first_fixing}.

\subsection{Higher order derivatives evaluated at $\Phi=0$}\label{ssec:higher_order}

To get conditions fixing the remaining coefficients, we need to turn to higher order differential operators. Unfortunately, obvious guesses like $\der{\Phi_{ab}}\der{\Phi_{ba}}$ yield functionals $O[\Phi]$ which are not invariant under $\Phi\rightarrow \Omega^\dagger \Phi\Omega$, and therefore do not give direct information about the coefficients we wish to fix. Instead, we have to consider such higher derivatives evaluated at $\Phi=0$. For example,
\begin{equation}
 \left.\der{\Phi_{ab}}\der{\Phi_{ba}} \de^{-\beta\tr(\Phi[L_i,[L_i,\Phi]])}\right|_{\Phi=0}=-4\beta\tr(L_i^2)~.
\end{equation}
Since the result is invariant under $\Phi\rightarrow \Omega^\dagger \Phi\Omega$, we also have
\begin{equation}
 \left.\der{\Phi_{ab}}\der{\Phi_{ba}}\int \dd \mu_\Omega~ \de^{-\beta\tr(\Omega^\dagger\Phi\Omega[L_i,[L_i,\Omega^\dagger\Phi\Omega]])}\right|_{\Phi=0}=-4\beta\tr(L_i^2)~.
\end{equation}
and finally
\begin{equation}\label{eq:second_fixing}
 \left.\der{\Phi_{ab}}\der{\Phi_{ba}} \de^{-\beta S_{\rm MT,kin}[\Phi]}\right|_{\Phi=0}=-4\beta\tr(L_i^2)~.
\end{equation}

Now to fix all the coefficients corresponding to a partition $\pi$ of an even integer $q$, it suffices to consider higher differential operators of order $q$ corresponding to $\pi$. For example, the multitrace terms at order four in the matrix $\Phi$,
\begin{equation}
\begin{aligned}
 &a_4\Phi_{ab}\Phi_{bc}\Phi_{cd}\Phi_{da}~,~~~&&a_{3,1}\Phi_{ab}\Phi_{bc}\Phi_{ca}\Phi_{dd}~,~~~&&a_{2,2}\Phi_{ab}\Phi_{ba}\Phi_{cd}\Phi_{dc}~,\\
 &a_{2,1,1}\Phi_{ab}\Phi_{ba}\Phi_{cc}\Phi_{dd}~,~~~&&a_{1,1,1,1}\Phi_{aa}\Phi_{bb}\Phi_{cc}\Phi_{dd}~,
\end{aligned}
\end{equation}
are fixed by considering the corresponding differential operators
\begin{equation}
\begin{aligned}
 &\der{\Phi_{ab}}\der{\Phi_{bc}}\der{\Phi_{cd}}\der{\Phi_{da}}~,~~~&&\der{\Phi_{ab}}\der{\Phi_{bc}}\der{\Phi_{ca}}\der{\Phi_{dd}}~,~~~&&\der{\Phi_{ab}}\der{\Phi_{ba}}\der{\Phi_{cd}}\der{\Phi_{dc}}~,\\
 &\der{\Phi_{ab}}\der{\Phi_{ba}}\der{\Phi_{cc}}\der{\Phi_{dd}}~,~~~&&\der{\Phi_{aa}}\der{\Phi_{bb}}\der{\Phi_{cc}}\der{\Phi_{dd}}~,
\end{aligned}
\end{equation}
if we know the lower order coefficients corresponding to partitions of all even integers smaller than $q$.

This can be readily seen as follows. Consider the partitions of an even integer $q$ in some chosen order and label the corresponding index contractions by a multiindex $I$. If we know the vector $\left.\dpar_I \de^{-\beta\tr(\Phi[L_i,[L_i,\Phi]])}\right|_{\Phi=0}$ and the matrix $M_{IJ}:=\dpar_I\Phi_J$ consisting of the higher derivatives of the multitraces, we can determine the coefficients corresponding to partitions of $q$ by inverting $M$. It is not hard to see that $M_{IJ}$ is invertible. First, note that the action of $\dpar_I$ on any $\Phi_J$ produces a string of fully contracted Kronecker deltas, which in turn will produce powers of $N$. If $I$ and $J$ are equal then the deltas will collapse to $N^n$ and lower powers of $N$ otherwise. For example,
\begin{equation}
 \der{\Phi_{a_1b_1}}\der{\Phi_{b_1c_1}}\der{\Phi_{c_1d_1}}\der{\Phi_{d_1a_1}}\Phi_{a_2b_2}\Phi_{b_2c_2}\Phi_{c_2d_2}\Phi_{d_2a_2}=\delta_{a_1a_2}^2\delta_{b_1b_2}^2\delta_{c_1c_2}^2\delta_{d_1d_2}^2=N^4~,
\end{equation}
while
\begin{equation}
 \begin{aligned}
 \der{\Phi_{a_1b_1}}\der{\Phi_{b_1c_1}}\der{\Phi_{c_1d_1}}\der{\Phi_{d_1a_1}}\Phi_{a_2a_2}\Phi_{b_2b_2}\Phi_{c_2c_2}\Phi_{d_2d_2}&=\\
 \delta_{a_1a_2}\delta_{b_1a_2}\delta_{b_1b_2}\delta_{c_1b_2}&\delta_{c_1c_2}\delta_{d_1c_2}\delta_{d_1d_2}\delta_{a_1d_2}=N~.
\end{aligned}
\end{equation}
Therefore the product of the diagonal elements of $M_{IJ}$ is dominant in $N$ which implies that for $N$ large enough, the determinant of $M_{IJ}$ is non-vanishing. Altogether, $M$ is invertible and determines the coefficients arising from partitions of $q$.

Note that this procedure will produce coefficients $a_\pi$ that are of homogeneous order $q/2$ in $\beta$, where $\pi$ is again a partition of the even integer $q$. Each total order in the field therefore corresponds to an order in the inverse temperature $\beta$. 

We can now combine this method with that of the previous section. That is, we can restrict ourselves to a sub-matrix of $M_{IJ}$, where the multiindices $I$ and $J$ run over partitions of $q$ without any parts of size 1. This significantly simplifies the computations at each order.

\section{The multitrace matrix model}

We now compute the multitrace matrix model, limiting ourselves to fourth order in the inverse temperature $\beta$. At this order, we can perform nontrivial checks against the results of \cite{Polychronakos:2013nca}.

\subsection{Remaining calculations}

To fully compute the multitrace action, it remains to calculate various traces over products of the $L_i$. For this, it is useful to have a basis $\tau_\mu$, $\mu=1,\ldots,(n+1)^2$ of $\au(n+1)$ satisfying
\begin{equation}
 \tr(\tau_\mu\tau_\nu)=\delta_{\mu\nu}\eand \tau_\mu^{\alpha\beta}\tau_\mu^{\gamma\delta}=\delta^{\alpha\delta}\delta^{\beta\gamma}~.
\end{equation}
We then readily compute the sum over the eigenvalues of the second Casimir to be
\begin{equation}
 \Sigma_1:=\tr(\tau_\mu C_2\acton \tau_\mu)=2\tr(L_iL_i)N~.
\end{equation}
We can demand that the $L_i$ are orthogonal, which yields
\begin{equation}
 \tr(L_iL_j)=\frac{\Sigma_1}{2N}\frac{1}{(n+1)^2-1}\delta_{ij}~.
\end{equation}
By considering the $L_i$ in the fundamental representation, we derive 
\begin{equation}
 f_{ijk}f_{ij\ell}=2 (n+1)\delta_{k\ell}~.
\end{equation}
Using these relations, we compute
\begin{equation}
 \tr(L_iL_jL_iL_j)=\frac{\Sigma_1^2}{4N^3}-(n+1)\frac{\Sigma_1}{2N}
\end{equation}
as well as
\begin{equation}
\begin{aligned}
 \tr(L_iL_jL_kL_jL_iL_k)&=\frac{\Sigma_1(-2(1+n)N^2+\Sigma_1)^2}{8N^5}~,\\
 \tr(L_iL_jL_kL_iL_jL_k)&=\frac{\Sigma_1(8(1+n)^2N^4-6(1+n)N^2\Sigma_1+\Sigma_1^2)}{8N^5}~.
\end{aligned}
\end{equation}

Moreover, the sum over the eigenvalues of the second Casimir cubed is
\begin{equation}
 \Sigma_3:=\tr(\tau_\mu C_2\acton C_2\acton C_2\acton \tau_\mu)=\frac{(3+2n+n^2)\Sigma_1^3}{n(2+n)N^4}-8\tr(L_iL_jL_k)\tr(L_kL_jL_i)~.
\end{equation}
and we have
\begin{equation}
\begin{aligned}
 \tr(L_iL_jL_k)\tr(L_kL_jL_i)&=-\tfrac{1}{8}\Sigma_3-\frac{(3+2n+n^2)\Sigma_1^3}{n(2+n)N^4}~,\\
 \tr(L_iL_jL_k)\tr(L_iL_jL_k)&=-\tfrac{1}{8}\Sigma_3-\frac{(3+2n+n^2)\Sigma_1^3}{n(2+n)N^4}-\frac{(1+n)\Sigma_1^2}{4n(2+n)N^2}~.
\end{aligned}
\end{equation}
Analogously, we compute the traces over terms of eighth order in the $L_i$. Since these expressions are very involved, we refrain from giving them explicitly.

\subsection{Multitrace action}

Combining the techniques of the previous section with the results above, we obtain the following rewriting:
\begin{equation}
 \int \dd \mu_{\Omega}~\de^{-\beta \tr(\Phi C_2\acton \Phi)}=\int \dd \mu_{\Omega}~\de^{-\beta S_{\rm MT,kin}[\Phi]}
\end{equation}
with
\begin{equation}
 \begin{aligned}
  S_{\rm MT,kin}[\Phi]=&a_2\left(\tr(\Phi^2)-\frac{1}{N}\tr(\Phi)^2\right)+\\
  &+a_4\left(\tr(\Phi^4)-\frac{4\tr(\Phi^3)\tr(\Phi)}{N}+\frac{6\tr(\Phi^2)\tr(\Phi)^2}{N^2}-\frac{3\tr(\Phi)^4}{N^3}\right)+\\
  &+a_{2,2}\left(\tr(\Phi^2)-\frac{\tr(\Phi)^2}{N}\right)^2+\ldots\\
  =&\sum_{\pi} a_{\pi} s(\pi)=\sum_{\pi} a_{\pi} s(\pi_1)s(\pi_2)\ldots s(\pi_k)~.\\
 \end{aligned}
\end{equation}
Here, the sum runs over partitions $\pi=\pi_1+\pi_2+\ldots+\pi_k$ of even integers not containing parts 1. The lowest four orders are governed by the following expressions:
\begin{equation}\label{eq:s-pi-1}
 \begin{aligned}
s(2)=&\tr(\Phi^2)-\frac{\tr(\Phi)^2}{N}~,\\
s(3)=&\tr(\Phi^3)-\frac{3\tr(\Phi)\tr(\Phi^2)}{N}+\frac{2\tr(\Phi)^3}{N^2}~,\\
s(4)=&\tr(\Phi^4)-\frac{4\tr(\Phi)\tr(\Phi^3)}{N}+\frac{6\tr(\Phi^2)\tr(\Phi)^2}{N^2}-\frac{3\tr(\Phi)^4}{N^3}~,\\
s(5)=&\tr(\Phi^5)-\frac{5\tr(\Phi^4)\tr(\Phi)}{N}+\frac{10\tr(\Phi^3)\tr(\Phi)^2}{N^2}-\frac{10\tr(\Phi^2)\tr(\Phi)^3}{N^3}+\\&+\frac{4\tr(\Phi)^5}{N^4}~,\\\
s(6)=&\tr(\Phi^6)-\frac{6\tr(\Phi^5)\tr(\Phi)}{N}+\frac{15\tr(\Phi^4)\tr(\Phi)^2}{N^2}-\frac{20\tr(\Phi^3)\tr(\Phi)^3}{N^3}+\\&+\frac{15\tr(\Phi^2)\tr(\Phi^4)}{N^4}-\frac{5\tr(\Phi^6)}{N^5}~,\\
s(8)=&\tr(\Phi^8)-\frac{8\tr(\Phi^7)\tr(\Phi)}{N}+\frac{28\tr(\Phi^6)\tr(\Phi)^2}{N^2}-\frac{56\tr(\Phi^5)\tr(\Phi)^3}{N^3}+\\&+\frac{70\tr(\Phi^4)\tr(\Phi)^4}{N^4}-\frac{56\tr(\Phi^3)\tr(\Phi)^5}{N^5}+\frac{28\tr(\Phi^2)\tr(\Phi)^6}{N^6}-\frac{7\tr(\Phi^8)}{N^7}~.
 \end{aligned}
\end{equation}
The general pattern for arbitrary $n\geq 2$, which is a result of relation \eqref{eq:first_fixing}, is
\begin{equation}\label{eq:s-pi-2}
 s(n)=\binom{n}{0}\tr(\Phi^n)+\sum_{k=1}^{n-2}(-1)^k\binom{n}{k}\frac{\tr(\Phi^{n-k})\tr(\Phi^k)}{N^k}+(-1)^{n-1}(n-1)\frac{\tr(\Phi^n)}{N^{n-1}}~.
\end{equation}

For the coefficients $a_\pi$, we obtain:
\begin{equation}\label{eq:intermediate_1}
\begin{aligned}
a_2=&\frac{\Sigma_1}{N^2-1}~,\\
a_4=&\frac{\beta  \Sigma_1 \left(2 n (1+n) (2+n) N^2 \left(1+N^2\right)-\left(6-4 N^2+n (2+n) \left(6+N^2\right)\right) \Sigma_1\right)}{n (2+n) N^3 \left(-36+N^2 \left(-7+N^2\right)^2\right)}~,\\
a_{2,2}=&\frac{\beta  \Sigma_1 \left(-2 n (1+n) (2+n) N^2 \left(3-5 N^2+2 N^4\right)\right)}{n (2+n) N^4 \left(-1+N^2\right)^2 \left(36-13 N^2+N^4\right)}+\\
&+\frac{\beta  \Sigma_1^2 \left(18 (1+n)^2-3 (8+3 n (2+n)) N^2+(7+3 n (2+n)) N^4-N^6\right)}{n (2+n) N^4 \left(-1+N^2\right)^2 \left(36-13 N^2+N^4\right)}~,
\end{aligned}
\end{equation}
where $\Sigma_1$ is again the trace over the eigenvalues of the quadratic Casimir $C_2$. Because the expressions for higher orders in $\beta$ get longer and more cumbersome to write down, let us restrict ourselves here to the case $n=1$, even though we can readily compute expressions for general $n$. For $n=1$, we have\footnote{Although the coefficients $a_\pi$ look simple for $n=1$, they are much more involved already for $n=2$.}
\begin{equation}
 a_2=N^2~,~~~a_4=\frac{N\beta}{3}~,~~~a_{2,2}=\beta\left(1-\frac{N^2}{3}\right)
\end{equation}
and
\begin{equation}
\begin{aligned}
 a_6&=0~,~~~&a_{4,2}&=-\frac{4\beta^2N(N^2+3)}{3(N^2-1)}~,\\
 a_{3,3}&=-\frac{4\beta^2N(N^4-12N^2-21)}{27(N^2-1)}~,~~~&a_{2,2,2}&=\frac{4\beta^2(5N^2+1)}{9(N^2-1)}~.
\end{aligned}
\end{equation}
At fourth order in $\beta$, we have the following terms for $n=1$.
\begin{equation}
\begin{aligned}
 a_8&=\frac{2}{135} N \left(13-10 N^2\right) \beta ^3~,\\
 a_{6,2}&=-\frac{8 \left(2 N^4+171 N^2+187\right) \beta ^3}{135 \left(N^2-1\right)}~,\\
 a_{5,3}&=\frac{16 \left(17 N^4-1525 N^2-940\right) \beta ^3}{135 \left(N^4-5 N^2+4\right)}~,\\
 a_{4,4}&=\frac{2 \left(-5 N^8+58 N^6+228 N^4+9947 N^2+4892\right) \beta ^3}{135 \left(N^4-5 N^2+4\right)}~,\\
 a_{4,2,2}&=\frac{4 \left(10 N^8-63 N^6-921 N^4-8062 N^2-2304\right) \beta ^3}{135 N \left(N^4-5 N^2+4\right)}~,\\
 a_{3,3,2}&=\frac{16 \left(13 N^6+191 N^4+1752 N^2+384\right) \beta ^3}{135 N \left(N^4-5 N^2+4\right)}~,\\
 a_{2,2,2,2}&=\frac{2 \left(-17 N^6+256 N^4+1147 N^2+5094\right) \beta ^3}{135 \left(N^4-5 N^2+4\right)}~.
\end{aligned}
\end{equation}

As a consistency check, we can consider the model for fuzzy $\CPP^n$ at first quantization level $\ell=1$, for which $N=n+1$. Here, $\sEnd(\CH_\ell)$ contains the adjoint representation of $\asu(n+1)$, and the kinetic term can be evaluated explicitly using the Fierz identity 
\begin{equation}\label{eq:FierzIdentitysun}
 L_i^{\alpha\beta}L_i^{\gamma\delta}=\delta^{\alpha\delta}\delta^{\beta\gamma}-\frac{1}{n+1}\delta^{\alpha\beta}\delta^{\gamma\delta}
\end{equation}
in this representation. We find that
\begin{equation}\label{eq:toymodelKineticTerm}
 \tr(\Phi C_2\Phi)= \frac{\Sigma_1}{N^2-1}\left(\tr(\Phi^2)-\frac{1}{N}\tr(\Phi)\tr(\Phi)\right)~.
\end{equation}
Starting from our action above, it is now a straightforward exercise to put $N=n+1$. Note that at first quantization level $\ell=1$,
\begin{equation}
 \Sigma_1=2n(1+n)(2+n)~,~~~\Sigma_2=4n(1+n)^2(2+n)~,~~~\Sigma_3=8n(1+n)^3(2+n)~,
\end{equation}
where $\Sigma_i$ is the sum over the $i$-th powers of the eigenvalues of the quadratic Casimir $C_2$. 

The terms $a_\pi s(\pi)$ for $\pi\neq 2$ now cancel as follows: For $n$ small compared to the integer partitioned by $\pi$, the expressions $s(\pi)$ vanish due to Newton's identity. For large $n$, the prefactors $a_\pi$ themselves vanish. Altogether, the kinetic part of the multitrace models collapses to \eqref{eq:toymodelKineticTerm}.

\subsection{The large $N$ limit}

Having rewritten scalar field theory on fuzzy complex projective space as a multitrace matrix model, the next step in computing the partition function of this model is to integrate out the angular degrees of freedom corresponding to zero modes. As the multitrace matrix model is invariant under the transformation $\Phi\rightarrow \Omega^\dagger \Phi\Omega$ for unitary $\Omega$, we can readily diagonalize $\Phi$ and integrate over the Haar measure. Taking the large $N$ limit, we can apply a saddle point approximation to compute the eigenvalue densities, which capture e.g.\ the phase diagram of fuzzy scalar field theory. The latter computations are readily performed, cf.\ \cite{Saemann:2010bw}, and we refrain from going into further details. However, we would like to describe the large $N$ limit of the action and to study the remainder term $R$ appearing in the form of the multitrace action \eqref{eq:conjectured_action} given in \cite{Polychronakos:2013nca}.

For brevity, let us restrict to the case of the fuzzy sphere. The large $N$ limit is a transition from finitely many to infinitely many degrees of freedom. As usual in quantum field theory, such a limit needs to be accompanied by a renormalization prescription, i.e.\ by scalings for all the fields and coupling constants. The multiscaling limit is here determined by two constraints. First, we require at least some parts of the kinetic term to survive the large scaling limit. Second, the total scaling of dominant terms in the action should be $N^2$ to match the scaling of the exponentiated Vandermonde determinant. As derived in \cite{Saemann:2010bw}, this implies that the field $\Phi$ is rescaled by a factor $N^{-1/4}$ and $\beta$ is rescaled by $N^{-1/2}$. The parameters $r$ and $g$ appearing in the potential are rescaled by factors $N^2$ and $N^{5/2}$. Moreover, in the transition from traces to integrals over eigenvalue densities, each trace comes with an additional factor of $N$.

The scaling of the traces by an additional factor of $N$ ensures that the $s(\pi)$, from which the action is built as $\sum_\pi a_\pi s(\pi)$, scale homogeneously. This is readily seen from equations \eqref{eq:s-pi-1} and \eqref{eq:s-pi-2}. We can therefore restrict our attention to those $a_\pi$ which are dominant in $N$ at each order in $\beta$. These are $a_{2}$, $a_{2,2}$, $a_{3,3}$, $a_{4,4}$, $a_{4,2,2}$ and $a_{2,2,2,2}$, which all scale homogeneously, and we have
\begin{equation}\label{eq:action-large-N-limit}
\begin{aligned}
 \lim_{N\rightarrow \infty} S_{\rm MT,kin}=N^2\Big(&\tilde s(2)-\frac{\tilde{\beta}}{3}\tilde s(2,2)-\frac{4\tilde \beta^2}{27}\tilde s(3,3)+\\&-\frac{10\tilde \beta^3}{135} \tilde s(4,4)+\frac{40\tilde \beta^3}{135}\tilde s(4,2,2)-\frac{34\tilde \beta^3}{135}\tilde s(2,2,2,2)+\ldots\Big)~.
\end{aligned}
\end{equation}
Here, $\tilde s(\pi)$ are the expressions $s(\pi)$ with traces now replaced by integrals over eigenvalue densities and $\tilde \beta$ is the renormalization of $\beta$. The lowest three orders in $\beta$ of this large $N$ limit perfectly agree with those found by group theoretic methods in \cite{Saemann:2010bw}.

We are now in a position to compare our results with those of \cite{Polychronakos:2013nca,Tekel:2014bta,Nair:2011ux} for the large $N$ limit. First of all, the relations \eqref{eq:intermediate_1} turn precisely into \cite[eq.\ (3.6)]{Tekel:2014bta} in the limit of large matrix size when taking into account the different conventions. 

Next, let us consider the constraints on the effective multitrace action obtained in \cite{Polychronakos:2013nca}. Recall that the eigenvalue distribution in the large $N$ limit is a renormalized Wigner semicircle and this limits the possible terms in the effective action to those that do not affect this shape. In particular, we have to lowest order that \cite{Polychronakos:2013nca}
\begin{equation}
 \tilde a_4=0~,~~~\tilde a_6=\tilde a_{4,2}=0~,~~~\tilde a_8=\tilde a_{6,2}=0~,~~~4\tilde a_{4,4}+\tilde a_{4,2,2}=0~,
\end{equation}
where $\tilde a_\pi$ are the coefficients of $\tilde s(\pi)$ appearing in \eqref{eq:action-large-N-limit}. These conditions are clearly satisfied by the terms appearing in \eqref{eq:action-large-N-limit}.

Moreover, the following form of the effective multitrace action has been derived in \cite{Polychronakos:2013nca}:
\begin{equation}
 \tilde \beta S_{\rm MT,kin}=\frac{N^2}{2}\left(\frac{\tilde s_2}{2}-\frac{\tilde s_2^2}{24}+\frac{\tilde s_2^4}{2880}+\ldots\right)+b_1(\tilde s_4-2\tilde s_2^2)^2+\ldots~.
\end{equation}
This is simply a perturbative expansion of \eqref{eq:conjectured_action}, with $R$ being the term proportional to $b_1$. With $\tilde \beta=\frac{1}{4}$ and $b_1=-\frac{10}{135}$, this agrees\footnote{Recall that terms containing traces of odd powers of the $\Phi$ were dropped in \cite{Polychronakos:2013nca}, as it was assumed that the vacuum is symmetric.} with \eqref{eq:action-large-N-limit}. 

Altogether we thus get a nice confirmation of the validity of our calculation. Recall that there was an implicit hope that $b_1=0$ in \cite{Polychronakos:2013nca}; our results, however, show that this is unfortunately not the case.

As a final remark, note that one might be tempted to try to derive the shape of the higher dominant terms in \eqref{eq:action-large-N-limit} analytically. Particularly suggestive is the fact that the matrix $M_{IJ}$ governing the bootstrapping conditions in section \ref{ssec:higher_order} necessarily becomes diagonal in the large $N$ limit. A brief examination of the third and fourth orders in $\beta$, however, shows that cancellations are quite involved and a truncated large $N$ analysis is not sufficient.

\section{Conclusions}

In this paper, we presented a new method of obtaining a perturbative high-temperature expansion for the action of scalar field theory on fuzzy complex projective spaces. This expansion yields a hermitian matrix model involving multitrace terms which can be treated by many of the usual matrix model techniques. The expansion can then be used to compute expectation values of operators invariant under $\Phi\rightarrow \Omega^\dagger \Phi \Omega$ for unitary matrices $\Omega$.

Our approach is very general. In principle, it is applicable to all scalar field theories on noncommutative spaces which are implicitly hermitian matrix models. The fixed external matrices appearing in their kinetic terms are again rewritten into very manageable multitrace expressions.

Although our method still requires some involved algebraic manipulations that are better left to computer algebra programs, it is certainly simpler and more powerful than the group-theoretic approach presented in \cite{Saemann:2010bw}. It is also more robust and more easily implemented algorithmically.

The method itself is based on a bootstrapping approach. We derived sufficiently many constraints by considering the action of differential operators on the exponentiated kinetic term to fix its high-temperature expansion. In the case of fuzzy scalar field theories, we first used a shift-symmetry of the kinetic term to determine simple analytical expressions for most of the coefficients at each order of the high-temperature expansion. Going beyond the symmetries of the model, we then computed the remaining coefficients up to fourth order in the inverse temperature $\beta$.

Our computation reproduced and extended the perturbative expansion of \cite{Saemann:2010bw}. Moreover, it confirmed the shape of the high-temperature expansion in the large $N$ limit that was derived in \cite{Polychronakos:2013nca}. Unfortunately, a number of terms in our expansion show that the multitrace rewriting is more involved than implicitly hoped in \cite{Polychronakos:2013nca}.

It is hard to anticipate any further improvement on our results based on the bootstrapping approach since we exploited all obvious symmetries. In future research, we plan to apply our method to various other noncommutative scalar field theories and study their phase structure in detail.

\section*{Acknowledgements}
I would like to thank Juraj Tekel for useful comments on a first version of this paper. This work was partially supported by the Consolidated Grant ST/L000334/1 from the UK Science and Technology Facilities Council.

\appendices

\subsection{Summary of conventions and notation}

We are working with the quantization of $n$-dimensional complex projective space $\CPP^n$. Its algebra of isometries is $\asu(n+1)$ and has $(n+1)^2-1$ generators. The dimension of the Hilbert space underlying the quantization is denoted by $N$. The relation between $N$, the level of quantization $\ell\in \NN_0$ and the dimension $n$ of the complex projective space is 
\begin{equation}
 N=\frac{(n+\ell)!}{n!\ell!}~.
\end{equation}
We have an $N$-dimensional representation of the generators of $\asu(n+1)$, denoted by $L_i$, $i=1,\ldots,(n+1)^2-1$. These satisfy the algebra
\begin{equation}
 [L_i,L_j]=:\di f_{ijk}L_k~,
\end{equation}
where the $f_{ijk}$ are the structure constants of $\asu(n+1)$. The normalization of the matrices $L_i$ is fixed by demanding that the eigenvalues of the second Casimir $C_2$,
\begin{equation}
C_2\acton \Phi:=[L_i,[L_i,\Phi]]~,
\end{equation}
are $2k(k+n)$ for $k=0,\ldots,\ell$ with multiplicities
\begin{equation}
 \frac{n(2k+n)((k+n-1)!)^2}{(k!)^2(n!)^2}~.
\end{equation}
That is, the sum over the eigenvalues of $C_2$ is
\begin{equation}
 \Sigma_1=\sum_{k=0}^\ell 2k(k+n)\frac{n(2k+n)((k+n-1)!)^2}{(k!)^2(n!)^2}~,
\end{equation}
and inductively, one can show that
\begin{equation}
 \Sigma_1=\frac{2\ell(1+\ell)^2(2+\ell)^2\cdots(n+\ell)^2((n+1)+\ell)}{(n+1)!(n-1)!}~.
\end{equation}
The matrices $L_i$ now satisfy the equations
\begin{equation}
 \tr(L_i)=0~,~~~L_i^2=\frac{\Sigma_1}{2N^2} \unit\eand\tr(L_iL_j)=\frac{\Sigma_1}{2N}\frac{1}{(n+1)^2-1}\delta_{ij}~.
\end{equation}
Finally, we have the following identity for the structure constants:
\begin{equation}\label{eq:StructureConstantsIdentity}
 f_{ijk}f_{ij\ell}=2 (n+1)\delta_{k\ell}~.
\end{equation}



\end{document}